# Jamming Distance Dictates Colloidal Shear Thickening

Shravan Pradeep[1], Mohamad Nabizadeh[2], Alan R. Jacob[1], Safa Jamali[2], and Lilian C. Hsiao[1,*]

[1] Department of Chemical and Biomolecular Engineering, North Carolina State University,

Raleigh, North Carolina – 27695, USA

[2] Department of Mechanical and Industrial Engineering, Northeastern University, Boston,

Massachusetts – 02115, USA

[*] Corresponding author: lilian_hsiao@ncsu.edu

**We report experimental and computational observations of dynamic contact networks for colloidal suspensions undergoing shear thickening. The dense suspensions are comprised of sterically stabilized poly(methyl methacrylate) hard sphere colloids that are spherically symmetric and have varied surface roughness. Confocal rheometry and dissipative particle dynamics simulations show that the shear thickening strength $\beta$ scales exponentially with the scaled deficit contact number $z^*$ and the scaled jamming distance $\Delta\phi/\phi_{max}$. Rough colloids, which experience additional tangential and rolling constraints, require an average of 1.5 - 2 fewer particle contacts as compared to smooth colloids, in order to generate the same $\beta$. This is because the surface roughness enhances geometric friction in a way that the rough colloids do not experience a large change in the free volume near the jamming point. In contrast, smooth colloids must undergo significant reduction in the free volume to support an equivalent shear stress. The available free volume for different colloid roughness is related to the deficiency from the maximum number of nearest neighbors at jamming under shear. Our results further suggest that the force per contact is different for particles with different morphologies.**

Dense suspensions of colloidal particles with stochastic Brownian motion exhibit shear thickening under flow, a non-Newtonian behavior where the suspension viscosity $\eta$ increases mildly or strongly depending on the applied shear stresses $\sigma$ and particle volume fraction $\phi$. The ability to design the onset of shear thickening provides a unique advantage in the reversible tuning of material mechanics, which is of great interest in fields such as soft robotics, impact resistant fabrics, and liquid manufacturing [1-3]. However, the tunability in these systems currently remains at a rudimentary level of "on" or "off". For dense suspensions to truly advance technology, the level of control over shear thickening needs to become more deliberate and refined [4,5]. In this letter, we show that designing shear thickening strength is possible for a broad class of colloidal suspensions through a singular parameter: the distance to jamming.

A jammed material at $\phi_J$ is conventionally defined as a disordered particulate system that has developed a yield stress [6]. Shear thickening shares similarities to jamming in that the particles in a flowing suspension become so impeded by nearest neighbors that they require an increasing amount of stress to continue flowing [1,7]. The microstructural origin of shear thickening was first attributed to the formation of hydroclusters in the Stokesian Dynamics simulations [8]. Experiments later corroborated this observation [9], suggesting that the onset of shear thickening can be discussed through a single dimensionless parameter, Peclet number ($Pe_{sh} = 6\pi\eta a_{eff}^3 \dot{\gamma} / k_B T$), that represents the strength of hydrodynamic forces to thermal motion. More recently, simulations that incorporate explicit interparticle friction or particle roughness plus lubrication hydrodynamics were able to fully capture the large increase in viscosity that is characteristic of strong shear thickening [10,11]. An important result from these simulations is the appearance of space-spanning force chains and velocity correlations in shear thickened suspensions [12]. These force chains arise from any combination of $\sigma$- and $\phi$-based constraints including hydrodynamics, repulsion, adhesion, and solid contact friction [11,13,14]. Beyond a critical $Pe_{sh}$, the force chains proliferate and grow stronger, leading to stronger shear thickening [15].

Interestingly, conventional microstructural characterization techniques such as the radial distribution function [14] or scattering patterns in the velocity-gradient-vorticity planes [16] are not sensitive to differences between shear thickened states. This observation implies that the distance between particle surfaces, rather than the center-to-center distance, is related to the force chains. As $\phi \rightarrow \phi_J$ and $\sigma$ increases, conservation laws state that the contact distance between particles in a constant-volume suspension must decrease, leading to a greater number of contacts, except for volume expansion at the free boundary for a few DST dominated flows. To address a lack of experimental evidence of force chains in the literature, we focus on the details of the dynamic contact networks formed by dense colloidal suspensions in shear thickening flows.

We use the mean contact number $<z>$, a measure for the number of contacting nearest neighbors around particles, to quantify the suspension microstructure because of the strong correlation of $<z>$ with bulk mechanics [17]. The contact number at jamming, $z_J$, and $\phi_J$ are inextricably linked to the interparticle friction in dense packings. For instance, application of Maxwell's isostatic criterion to a frictionless hard sphere system at $\phi_J = 0.64$ reveals that $z_J = 6$. Incorporating an infinitely large sliding ($\mu_s$) and rolling ($\mu_r$) friction coefficient between colloids, further reduces $\phi_J$ and $z_J$ [18,19]. The sliding constraint $\mu_s$ is featured in several constitutive equations, particle simulations, and models that describe shear thickening as due to particles undergoing a stress-induced lubricated-to-frictional transition beyond an onset stress $\sigma^*$ [20-22]. Both $\mu_s$ and $\mu_r$ are thought to generate long-lasting force chains by reducing the rotational degree of freedom of particles in flow [19]. Experimental measurements support this idea by demonstrating that the rotational dynamics of shape-symmetric particles with protrusions deviate significantly from simulations of hard sphere suspensions [23-25]. While the interparticle friction may not always track with surface roughness because of complex tribological factors (*e.g.*: elastohydrodynamics [26,27]), in general, rougher particles have larger values of $\mu_s$ and $\mu_r$.

To investigate the role of friction on the contact microstructure of shear thickening colloidal suspensions, we use confocal rheometry experiments and dissipative particle dynamics (DPD) simulations to identify a quantitative link between the strength of thickening $\beta = \log(\Delta\eta)/\log(\Delta\sigma)$ and the distance from jamming ($\phi_{max}$ - $\phi$)$/\phi_{max} = \Delta\phi/\phi_{max}$ for smooth and rough colloids. Here, $\phi_{max}$ refers to the maximum jamming fraction for a disordered packing, where $\phi_{max}$ = $\phi_J$ ($\sigma$ = 0 Pa) is obtained from confocal microscopy performed on colloids that have undergone unperturbed sedimentation under gravitational stress for three months. We obtain $\beta$ using the average slope at the inflection points above $\sigma^*$ and before the high shear plateau. At $\phi_{max}$, the suspension is considered mechanically rigid and is not flowable at or beyond this volume fraction. The value of $\phi_{max}$ is verified independently within an experimental uncertainty of $\pm$ 5% by fitting the relative low-shear viscosity ($\eta_{r,low-shear}$) divergence to the form $\eta_{r,low-shear} = (1 - \phi/\phi_{max})^{-2}$. The value of $\phi_{max}$ is a key parameter in normalizing the jamming distance because it varies for colloids with different surface morphologies.

Earlier treatise on suspension rheology have seen the prolific use of smooth hard spheres, and only recently have poly(methyl methacrylate) (PMMA) and silica colloids with controlled surface roughness become widely available [28]. We hypothesize that there is a universal correlation between $\Delta\phi/\phi_{max}$, $\beta$, and $<z>$ for all suspensions exhibiting shear thickening. To reveal this relationship, we synthesize spherically symmetric and size-monodisperse PMMA microspheres with different levels of surface roughness [29]. These particles are sterically stabilized with poly(12-hydroxystearic acid) (PHSA) brushes of lengths 10 – 15 nm [30]. We prepare suspensions at $\phi < \phi_{max}$ by first centrifuging the stock suspension at gravitational Peclet number, $Pe_g$ = 1500 ($Pe_g$ = $4\pi a_{eff}^{d}\Delta\rho g/3k_BT$), and subsequently diluting the shear jammed sediments with known volumes of solvent. We obtain $\phi$ by imaging the fluorescent colloids with confocal laser scanning microscopy (CLSM, Leica SP8) and processing the 3D image volumes using a brightness-weighted centroid-based algorithm [31]. Separately, steady shear rheological measurements are performed using a stress-controlled rheometer (TA Instruments DHR-2) fitted with a 50-mm sandblasted cone-and-plate geometry.

Fig. 1 shows different rheological behavior of PMMA hard colloids with two types of morphology and similar effective swollen diameters $2a_{eff}$, smooth (S, $2a = 1.65$ μm $\pm 4\%$, Fig. 1a) and rocky (RK, $2a_{eff} = 1.49$ μm $\pm 6\%$, Fig. 1b). Two other morphologies are also studied: slightly rough (SR, $2a_{eff} = 1.82$ μm $\pm 5\%$), very rough (VR, $2a_{eff} = 1.47$ μm $\pm 6\%$). These steady shear flow curves describe the relative suspension viscosity ($\eta_r = \eta/\eta_s$, solvent viscosity $\eta_s = 0.012$ Pa·s) as a function of scaled $\sigma$. The dotted lines represent the two stress points at which we obtain $<z>$ values from dynamic packings: one at $\tilde{\sigma} = \tilde{\sigma}_{\beta=0} < \tilde{\sigma}^*$ and the second at $\tilde{\sigma} = \tilde{\sigma}_{\beta} > \tilde{\sigma}^*$, where the overhead $\sim$ represents the stress values scaled with a factor of $a_{eff}^3/k_BT$. In our flow curves, we choose $\tilde{\sigma}_{\beta=0} \approx 200$ at the low-shear viscosity plateau and $\tilde{\sigma}_{\beta} \approx 10^4$ at the shear thickening regime for contact microstructure studies, as shown in Fig. 1. At $\tilde{\sigma}_{\beta}$, where $\tilde{\sigma} > \tilde{\sigma}^*$, the steric and lubrication layers between the colloids gives way to the proliferation of interparticle contacts [21,32].

Regardless of the surface morphology, the suspensions transition from fully Newtonian flow at low $\sigma$ and $\phi$, to continuous shear thickening (CST, $\beta < 1$) at intermediate $\sigma$ and $\phi$, and finally to discontinuous shear thickening (DST, $\beta \geq 1$) at high $\sigma$ and when $\phi$ is close to $\phi_{max}$. Suspensions also exhibit a secondary plateau at the highest values of $\sigma$, commonly observed in shear thickened states in which the particles' motion is hindered by either frictional or hydrodynamic forces [11,14,19,21,33]. The onset of DST for smooth particle suspensions occurs at $\phi = 0.55$ (Fig. 1a), which is similar to the values reported earlier in the literature for colloids interacting with a short-range repulsive potential [34,35].

Our data show that $\Delta\phi/\phi_{max}$ predicts $\beta$ for different types of colloidal suspensions containing spherically symmetric particles. Fig. 2 shows that all colloidal suspensions obey the general scaling of the form, $\beta \sim \exp(-\Delta\phi/\phi_{max})$, where DST ($\beta \approx 1$) is present at $\Delta\phi/\phi_{max} \leq 0.1$ and CST ($\beta < 1$) is found at $\Delta\phi/\phi_{max} > 0.1$. The value of $\beta$ rapidly decreases at increasing $\Delta\phi/\phi_{max}$. Additional support for this correlation comes from $\beta$ and $\Delta\phi/\phi_{max}$ values extracted from a number of literature studies: both experiments and simulations [11,19,22,32-34,36-39]. This scaling has significant impact in the academic and industrial communities because it enables the *a priori* estimation of $\beta$ (a non-equilibrium parameter) using $\Delta\phi/\phi_{max}$ (an equilibrium parameter). The remarkable agreement between experiments and simulations from independent research groups suggests that there exists a direct link between the dynamic shear thickening microstructure of colloids and their respective equilibrium jamming distance. This link is more clearly illustrated using the dynamic $<z>$ values of shear thickening suspensions to understand how the contact networks are related to $\Delta\phi/\phi_{max}$.

To characterize the contact microstructure of dense suspensions at the large applied stresses used to induce shear thickening, we use a custom confocal rheometer setup (Fig. 3a), where a stress-controlled rheometer (Anton Paar MCR 502 WESP) is directly coupled to a CLSM (Leica SP8) similar to an earlier set up described in the literature [40]. Steady shear is applied to suspensions of smooth and rough colloids using a 20 mm parallel-plate top geometry and a glass coverslip at the bottom with thickness $0.16 - 0.19$ mm. The confocal rheometer is used to obtain 3D image volumes of dense suspensions undergoing steady shear at $\tilde{\sigma}_{\beta=0}$ and $\tilde{\sigma}_{\beta}$, as described in Fig. 1. Each stack of size 50 μm $\times$ 50 μm $\times$ 10 μm is imaged in under 5 s and contain ~ $10^4$ particles. The suspensions contain 5 wt% mixture of photopolymer and photoinitiator additives to rapidly arrest the suspensions with ultraviolet (UV) light within 1 s (Supplemental Material S4). To obtain the sheared microstructure, we hold the suspensions at constant stresses, at values marked in Fig. 1, for 150 s and shine UV light ($\lambda_{exc} = 405$ nm) before and immediately drop the applied stress to zero, thus locking in the sheared suspension microstructure. We perform three independent experiments and obtain image stacks from three different points in each sheared sample. All image stacks are imaged at least 15 μm above the coverslip to avoid wall effects.

The images obtained from the confocal rheometer experiments are supported using dissipative particle dynamics (DPD) simulations of bidisperse suspensions ($a$ and $1.1a$ in an equal volume ratio with total number of particles $N = 1000$) containing smooth and rough colloids closely representing the experimental system. The particle roughness is modeled by distributing asperities of lengthscale $0.1a$ on the surface of the smooth base spheres, using a scheme similar to simulations in [11,14] (Supplemental Material S5). The solvent is modeled through explicit soft particles and interact through conservative, dissipative, and random Brownian forces. We have shown previously that this modified DPD scheme recovers a full representation of hydrodynamics in the system, similar to a fully resolved Stokesian Dynamics formalism. To compare the data from simulations and experiments, we use the suspension systems with smooth and rough particles in simulations match the $\phi_{max}$ to suspensions with S and RK systems from the experiments, respectively. The goal is to link $\beta$ to $<z>$ to capture the contact networks responsible for the shear thickening phenomena.

Defining interparticle contact during shear thickening requires the use of two different contact criteria at $\tilde{\sigma} < \tilde{\sigma}*$ and at $\tilde{\sigma} \geq \tilde{\sigma}*$, because the particles undergo a transition from lubricated to frictional flow and the soft PHSA brush becomes compressed by the large applied stresses [36]. At $\tilde{\sigma} < \tilde{\sigma}*$, two particles are defined to be in hydrodynamic contact if the interparticle separation is equal or less than the uncertainties that include the PHSA brush length, size polydispersity, and surface roughness [29]. At $\tilde{\sigma} \geq \tilde{\sigma}*$, a frictional contact is defined by the average center-to-center distance between particles, $2a_{eff}$ as shown in Fig. 3b. In DPD simulations, interparticle contacts are defined similarly for all particles and their interactions with other asperities and base particles. Experimental results are in excellent agreement with the contact microstructure obtained from DPD simulations for smooth and rough particles: the $<z>$ values obtained from DPD simulations fall within the error limits of the $<z>$ values obtained from our experimental packings, as shown in Fig. 4a.

Visual Molecular Dynamics (VMD) renderings of the dynamic packings, at $\tilde{\sigma}_\beta$, from the experiments and simulations for suspensions containing smooth and rough particles at $\Delta\phi/\phi_{max} = 0.08$ and $\beta \approx 0.85$ are shown in Fig. 3c-d. The renderings show the presence of space-spanning contact networks in both experiments and simulations of shear thickening flows and provide a statistical view of how smooth and rough pack differently in non-equilibrium shear thickening flows. Particles are concentrated in the compressive flow axis, in agreement with previous neutron scattering studies on shear thickening suspensions [16,41]. A first step towards constructing a statistical mean-field description parameter of the contact microstructure formed in such networks would be possible by evaluating the relationship between the dynamic contact number at $\tilde{\sigma}_\beta$, $<z>_\beta$, and $\beta$ for suspensions at various $\phi$.

Fig. 4a shows the dynamic contact number, $<z>_\beta$, as a function of $\Delta\phi/\phi_{max}$ for sheared suspensions of smooth and rough colloids. The dashed lines in Fig. 4a indicate that the smooth colloids, on average, requires an additional of $1.5 - 2$ contacts to maintain the same $\beta$ as compared to the rough colloids. This could be a consequence of a difference in available free volume for different types of colloid roughness and is related to the deficiency from the maximum number of nearest neighbors at jamming under shear. The value of $<z>_\beta$ is a function of $\tilde{\sigma}_\beta$ because the external deformation imparts an additional non-equilibrium free energy that must be minimized for steady flow [42]. To normalize the spatial effect of interparticle contacts that stem from free volume differences, we define a parameter $z^*$ that captures the scaled contact deficit, where $z^* =$ $(z_{J,\beta} - <z>_\beta)/z_{J,\beta}$. Here, $z_{J,\beta}$ is the maximum possible contacts available at $\phi_{J,\beta}$, which is defined as the divergence of the viscosity at $\tilde{\sigma}_\beta$ and indicate the maximum flowable volume fraction at $\tilde{\sigma}_\beta$.

To estimate the shear-induced jamming point $\phi_{J,\beta}$ for suspensions of smooth and rough colloids, we invoke an argument that relates the divergence of $\eta_r$ to $(\phi_J - \phi)$ at a given $\sigma$, where $\phi_J = \phi_J(\sigma)$. Specifically, the low-shear and high-shear viscosities are expected to diverge at $\phi_{max}$ and a $\sigma$ dependent $\phi_J$, respectively, with an exponent of -2 [43]. By extension, this suggests that $\eta_r$ at intermediate $\sigma$ should also diverge to a corresponding stress-dependent quasi-jamming point, $\phi_{J,\beta} = \phi_{J,\beta}(\tilde{\sigma}_\beta)$ with the same exponent of -2. The inset in Fig. 4b shows the scaling of the form $\eta_r \sim (\phi_{J,\beta} - \phi)^{-2}$ where $\phi_{J,\beta} = 0.61$ and $0.51$ for smooth and rough colloids, respectively. The value of $z_{J,\beta}$ is then obtained by extrapolating $<z>_\beta$ at various $\phi$ to the respective $\phi_{J,\beta}$, where $z_{J,\beta}$ as $4.95 \pm 0.01$ and $3.25 \pm 0.01$ for smooth and rough colloids. Fig. 4b shows that the dynamic contact scaling takes the form $z^* \sim (\Delta\phi/\phi_{max})^\alpha$ with $\alpha = 0.95 \pm 0.07$. A similar scaling ($\alpha = 1.08$) had been observed in 2D simulations of soft frictionless particles that are repulsive [44]. The observed power-law correlation in Fig. 4b is statistically significant with a normalized chi-squared parameter $\overline{\chi}_0^2 = 2.12$ and $P < .005$ [45] .

In Fig. 1, following the dashed line corresponding to $\tilde{\sigma}_\beta$ vertically, an increase in $\phi$ is associated with an increase in $\beta$ as more contacts are made. At the same time $\Delta\phi/\phi_{max}$ and $z^*$ decreases forming more space-spanning contacts and force networks. For a given shear thickening $\sigma$, for each particle system, there exists a $\phi_J$ and corresponding $z_J$ beyond which there is no steady state flow. Thus, smooth and rough colloids with similar $2a_{eff}$ are characterized with different $\phi_{J,\beta}$ and $z_{J,\beta}$ at the same applied $\tilde{\sigma}_\beta$. In a constant volume rheological experiment restricted by the dimensions of experimental and simulation setup, the free volume available to rearrange under shear is greater for smooth colloids than that of the rough colloids, because smooth colloids can rotate freely with little hydrodynamic resistance. There are more spatial constraints imposed by the restricted rotational degree of freedom and the effective rolling friction of rough colloids, and this is in turn captured by the deficiency of nearest neighbors to the shear-jamming contact number, $z_{J,\beta}$. The universality in Fig. 4b shows that this physical mechanism for shear thickening holds for all types of suspensions and thus the parameter $z^*$, which is a contact network parameter that captures the distance to $z_{J,\beta}$, can be used as the manifestation of the modes of particle motion under shear.

The dynamic contact scaling $z^* \sim \Delta\phi/\phi_{max}$ (Fig. 4b) and correlation $\beta \sim \exp(-\Delta\phi/\phi_{max})$ (Fig. 2) can be combined to relate the sheared contact microstructure and the shear thickening strength as $\beta \sim \exp(-z^*)$. The results suggest that at a given $\beta$, because $<z>$ is different for suspensions of particles with different asperities, the force carried by each contact is different for particles of different morphologies. Earlier work on compressed hydrogel beads found that the average macroscopic force, $F$, scales with average dynamic contacts as $F \sim <z>$ [46]. To obtain the same change in suspension stress (or $\beta$), rough particles suspensions required, on average, fewer contacts compared to suspensions with smoother particles. In other words, for the same $F$ in our systems, $F/<z>$ for rough particle suspensions must be greater than that of smooth counterpart. We indirectly capture the force per contact through parameter $z^*$. Note that the contact networks found in this work would likely have different morphologies and properties from the force chains observed in previous studies [12].

As a suspension shear thickens, clusters and percolated networks of particle contacts break and reform, but our study has shown that a mean-field description using dynamic $<z>$ can connect $\beta$ and $\Delta\phi/\phi_{max}$. The dynamic contact scaling may break down at $\phi$ values close to $\phi_{max}$ ($\Delta\phi/\phi_{max} \leq$ $10^{-2}$) due to pronounced flow instabilities such as localized stress fluctuations, periodic density fluctuations in flow direction, large transient clusters in compressive direction, and stress bands along the dilatational direction [47-49]. The increase in uncertainty in $z*$ close to the jamming point could be due to these flow instabilities. Nonetheless, our study shows that the scaled jamming distance is a strong predictor for the shear thickening behavior of a broad class of suspensions.

Because force networks are likely coupled to the contact network and particle positions [50], future studies that analyze the transient network anisotropy could provide new insight as to how different types of frictionless and frictional particles carry load in flowing systems. Athermal suspensions [51] and shape-anisotropic colloids [52] have not been tested in this study, and it would be interesting to see if the proposed scaling between $\beta$, $\Delta\phi/\phi_{max}$, and $z*$ hold for these materials.

**Acknowledgements.** We thank John Brady, Ronald Larson, Jeffrey Morris, and Abhinendra Singh for discussions. This work is supported in part by the National Science Foundation (NSF CBET-1804462), the American Chemical Society Petroleum Research Fund (ACS-PRF #59208-DNI9), and North Carolina State University start-up funds.

**Author Information.** All correspondence should be addressed to L.C.H (lilian_hsiao@ncsu.edu).

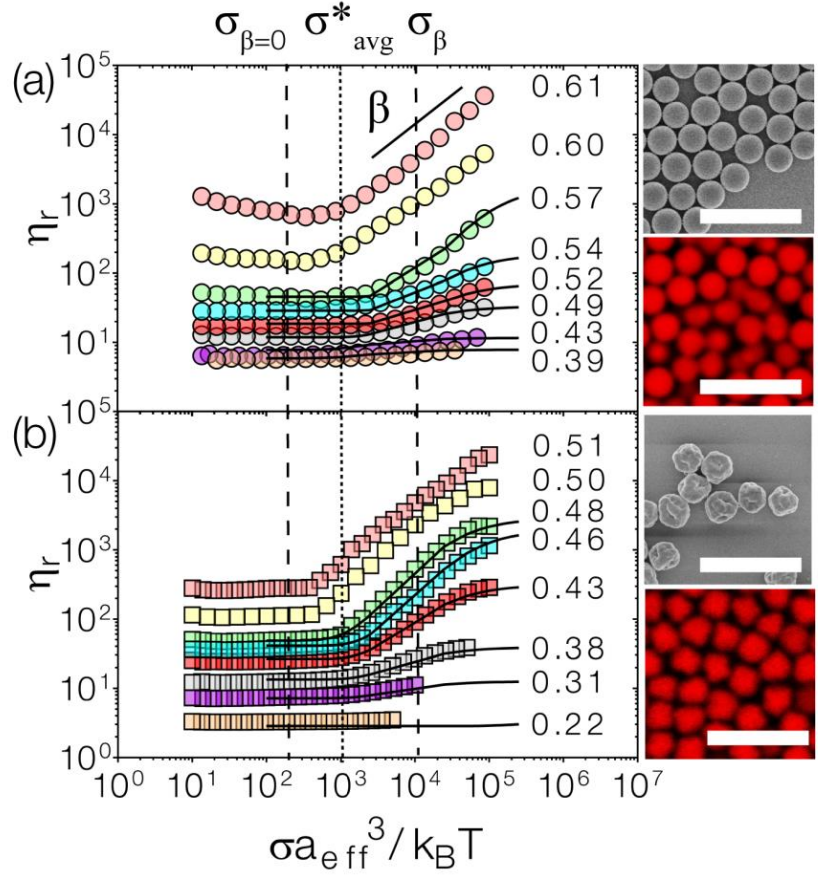

FIG. 1. Experimental rheology for suspensions of (a) smooth and (b) rough colloids. Flow curves represent $\eta_r$ plotted against $\sigma$ scaled by the effective particle radii and temperature. Numerical values next to each curve indicate $\phi$ (filled). Solid lines are fits with the WC theory. The vertical dashed lines represent the stresses below and above the onset stress (vertical dotted line) where we obtain the average contact number. Representative scanning electron micrographs and confocal micrographs of colloids are shown to the right side of respective flow curves. Scale: 5 μm.

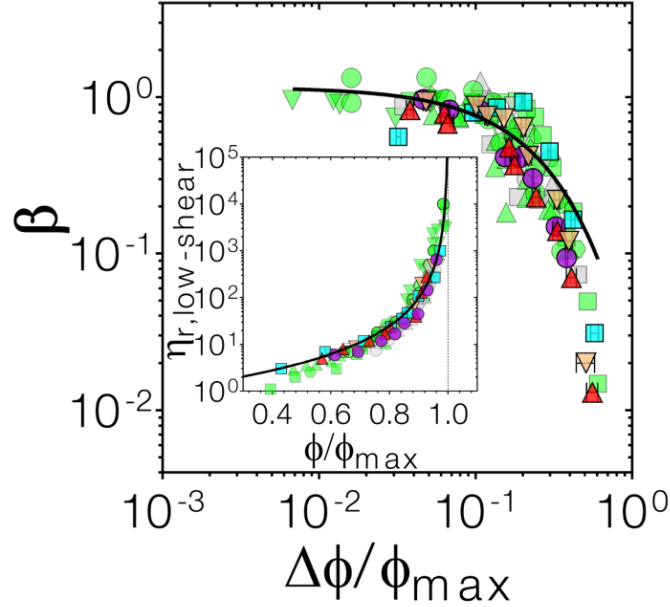

FIG. 2. Shear thickening strength as a function of jamming distance. Data from this work are shown for S (magenta circles), SR (red upper triangles),VR (coral lower triangles), and RK (cyan squares) colloids. Solid line indicates an emperical fit of the form: $\beta = \beta_0 \exp(\Delta\phi/\phi_{max} k)$ with $\beta_0 = 1.61 \pm 0.05$ and $k = -4.18 \pm 0.32$. Literature values from experimental colloidal studies are indicated by green symbols: smooth PMMA (circle) [37], rough PMMA (upper triangle) [33], smooth silica (square [40] and (hexagon) [36]), and rough silica (lower triangle [39] and diamond [34]). Literature values from simulations are indicated by grey symbols: colloids with surface asperities interacting via lubrication (square) [11], spheres with sliding friction (upper triangle) [22], spheres with sliding and rolling friction (circle) [19], and colloids interacting via sliding friction (lower triangle) [32]. Inset shows the fitting to the form: $\eta_r = (1-\phi/\phi_{max})^{-2}$ normalized for each particle $\phi_{max}$ values. Solid line represents the universal low-shear viscosity divergence.

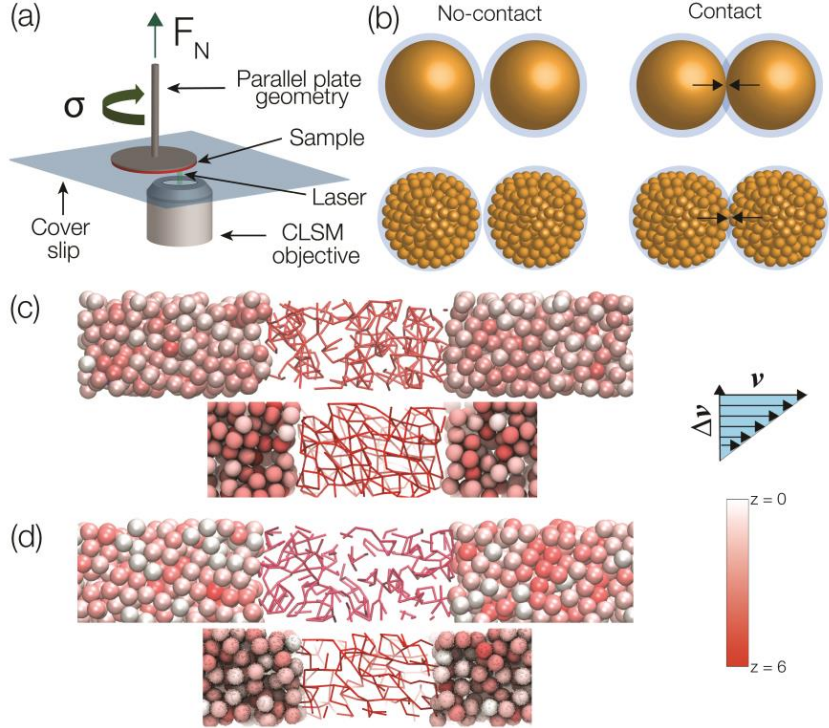

FIG. 3. (a) Confocal rheometer setup for imaging the shear-induced contact networks during flow measurements. (b) Contact criterion for interparticle contact in smooth (top row) and rough (bottom row). The light blue circle represents other lengthscales into consideration. (c,d) Contact networks of shear thickening suspensions at $\Delta\phi/\phi_{max} \approx 0.075$ and $\beta \approx 0.85$ for dense suspensions of (c) smooth and (d) rough colloids. For (c) and (d), the top panel are VMD reconstruction of the experimental microstructure and bottom panel are snapshots from simulations. The inset in the top panel represents the velocity-velocity gradient flow direction with respect to the contact networks shown in (c,d). The bottom inset shows the color panel for the contact number of the particles shown in (c,d).

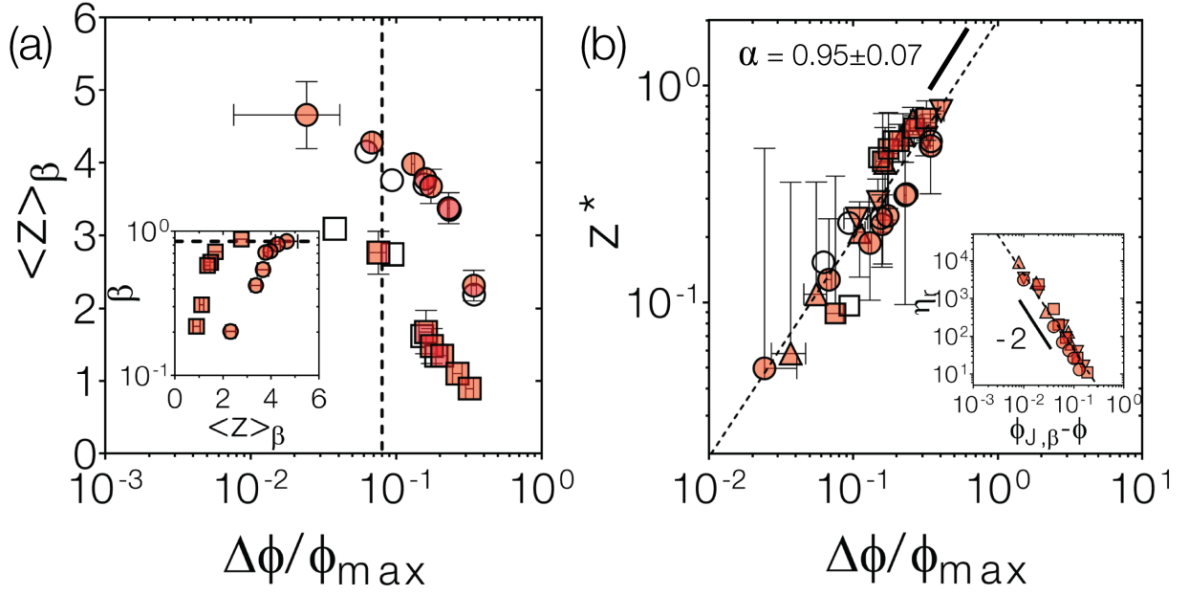

FIG. 4. (a) The change in $\langle z \rangle_\beta$ of smooth (circles) and rough (squares) colloids from experiments (filled) and simulations (unfilled) as a function of $\Delta\phi/\phi_{max}$. Inset shows the $\beta$ as a function of $\langle z \rangle_\beta$. Dashed lines in the main figure and the inset corresponds to the suspensions at $\Delta\phi/\phi_{max} \approx 0.075$ and $\beta \approx 0.85$ (b) The scaling $z^* \sim (\Delta\phi/\phi_{max})^\alpha$. Dashed line indicates the power law fit. Inset shows the scaling relation between the relative viscosity and unscaled jamming distance to test the fit, $\eta_r \sim (\phi_J - \phi)^{-2}$. Two additional types of rough particles: SR (upper triangles) and VR (lower triangles) are included here.

# Supplemental Material

# Jamming Distance Dictates Colloidal Shear Thickening

Shravan Pradeep[1], Mohamad Nabizadeh[2], Alan R. Jacob[1], Safa Jamali[2], and Lilian C. Hsiao[1,*]

**Affiliations:**

[1] Department of Chemical and Biomolecular Engineering, North Carolina State University, Raleigh, North Carolina – 27695, USA

[2] Department of Mechanical and Industrial Engineering, Northeastern University, Boston, Massachusetts – 02115, USA

[*] Corresponding author: lilian_hsiao@ncsu.edu

**Table of Contents**

## S.1. Flow fit curves from mean-field theory

Predictions of the suspension viscosity from the Wyart-Cates (WC) theory [1] are fitted to the experimental data as solid lines in Fig. 1. The original WC model attributes shear thickening to a change in the contact microstructure. First, in the low-$\sigma$ regime, hard spheres are frictionless and display a viscosity divergence at $\phi_{max} \approx 0.64$, so long as the particles remain separated in flow. An increasing subpopulation of particles then undergo a lubricated-to-contact transition at $\sigma \geq \sigma^*$, where the fraction of contacting particles is modeled by the sigmoidal form $f(\sigma) = \exp[-(\sigma^*/\sigma)^\gamma]$ and $\gamma$ is between 0.5 and 1.0 [2,3]. Finally, in the high-$\sigma$ regime, most of the particles interact through frictional mechanics and the suspension diverges in viscosity at $\phi_{J,f} \approx 0.58$, where the subscript denotes the frictional shear-jammed state. Under shear, these packings become more frictional and jam at even lower volume fractions ($\phi_{max,\mu} < \phi_{max}$, where $\mu$ is the interparticle friction coefficient). A spherically symmetric colloid with anisotropic surface morphology imparts different values of $\phi_{max}$ and $\phi_{J,f}$ as opposed to a smooth, frictionless hard sphere. From a purely granular perspective, dense suspensions shear thicken because of a transition between a packing with viscosity that diverges at $\phi_0$ to a new packing that diverges at $\phi_m$. The WC model generates flow curves through the following empirical relations:

$$\sigma = \lambda \left( \phi_J \left( f \right) - \phi \right)^{-2} \dot{\gamma} \quad \text{...........................(1)}$$

$$\phi_J \left( f \right) = \phi_{max} \left( 1 - f \right) + \phi_{J,f} f \quad \text{..................(2)}$$

The values for $\phi_{max}$ are 0.64 and 0.54 for smooth and rough particles, respectively. These values are obtained independently from low-shear viscosity divergence and maximum packing fraction from sedimentation experiments performed over three months. The estimated values for $\phi_{J,f}$ are performed using a similar viscosity divergence but at the shear-thickening plateau. Since we do not observe plateau for smooth particles for our system, we assumed $\phi_{J,f}$ as 0.58 from the original model. Our system gives the high-shear viscosity divergence at $\phi_{J,f} = 0.49$. The values of $f$ are varied from 0 to 1 as described by sigmoidal model above. The value of $\lambda$ in the above equation is an empirical fitting parameter.

## S.2. Rheology flow curves of SR and VR type colloids:

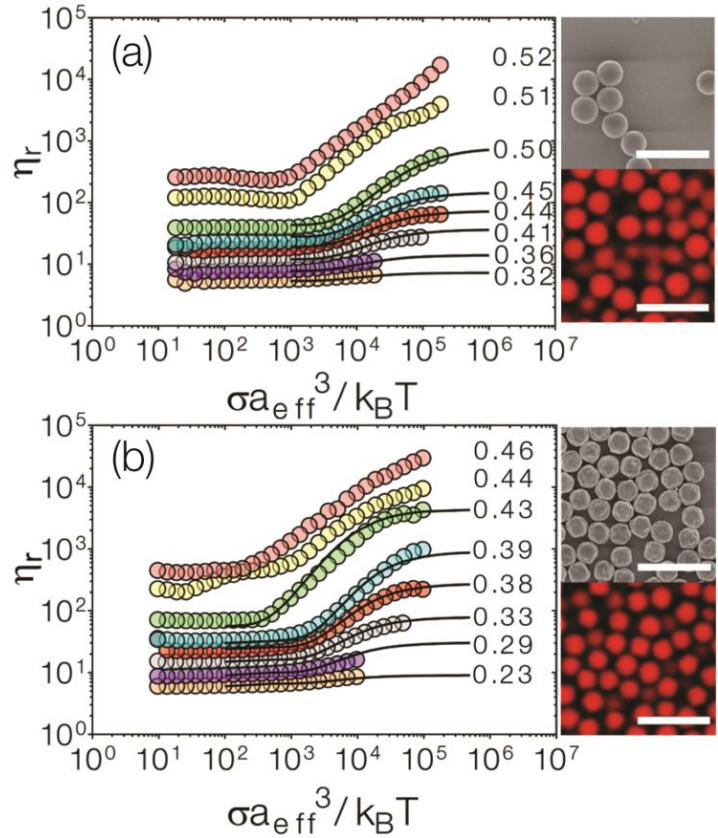

**Figure S1** | Viscosity-stress curves for suspensions of (a) SR and (B) VR colloids. Solid lines are fits with the Wyart-Cates model. The numbers on the side of each curve represent the average estimated $\phi$ of the respective suspension. Inset: Representative scanning electron micrographs and confocal laser scanning micrographs of colloids. Scale bars = 5 µm.

## S.3. Estimating jamming points ($\phi_{max}$) for colloidal suspensions

Superimposing our experimental values of $\phi_{max}$ and the corresponding $<z>$ onto the jammed simulation data of Singh *et al.* [4] suggests that our rough colloids can be modeled as particles with significant $\mu_s$ and $\mu_r$ (Figure S2) values.

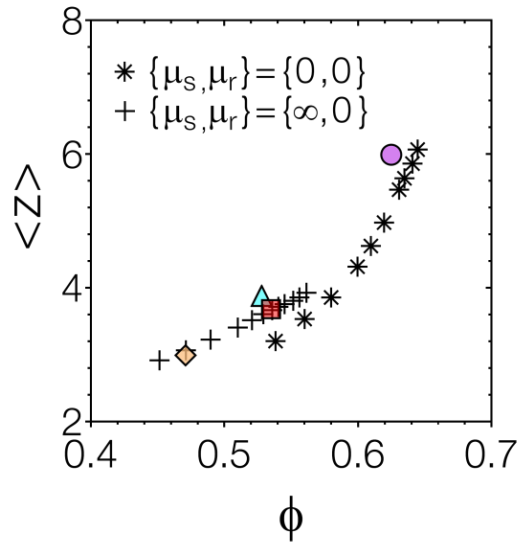

**Figure S2** | Experimental contact number $<z>$ for different suspensions plotted against $\phi$. Data are shown for S(magenta circles), SR (orange diamonds), VR (coral squares), and RK colloids (cyan triangles). Simulation data from particles interacting via short-range hydrodynamics, repulsion, and sliding/rolling friction are overlaid in the plot ($\mu_s$ = sliding friction, $\mu_r$ = rolling friction).

## S.4. Immobilization and visualization of sheared microstructure:

Rheology is performed using a 20 mm parallel-plate geometry and cover slip with thickness #1.5. Suspension samples are incorporated with 5 wt% photocrosslinker mixture (2 wt% photoinitiator phosphine oxide, 8 wt% photopolymer trimethylolpropane triacrylate-TMPTA, and solvent diisooctyl phthalate). Good agreement in the radial distribution function, $g(r)$, for suspension with and without the photocrosslinker (Figure S3(a)) shows that suspension doped with photocrosslinker mixture retain their hard sphere behavior. We estimate the mean-squared displacement $<\Delta r^2(t)>$ of a representative photopolymerized suspension. As shown in Figure S3(b), the noise floor is estimated to be 0.0012 $\mu m^2$ which is $\approx$ 0.1% of $(2a)^2$. This shows that the vibrations of the confocal rheometer is not significant enough to affect the particle centroid identification in the photopolymerized sample and the $<z>$ obtained from image processing routine.

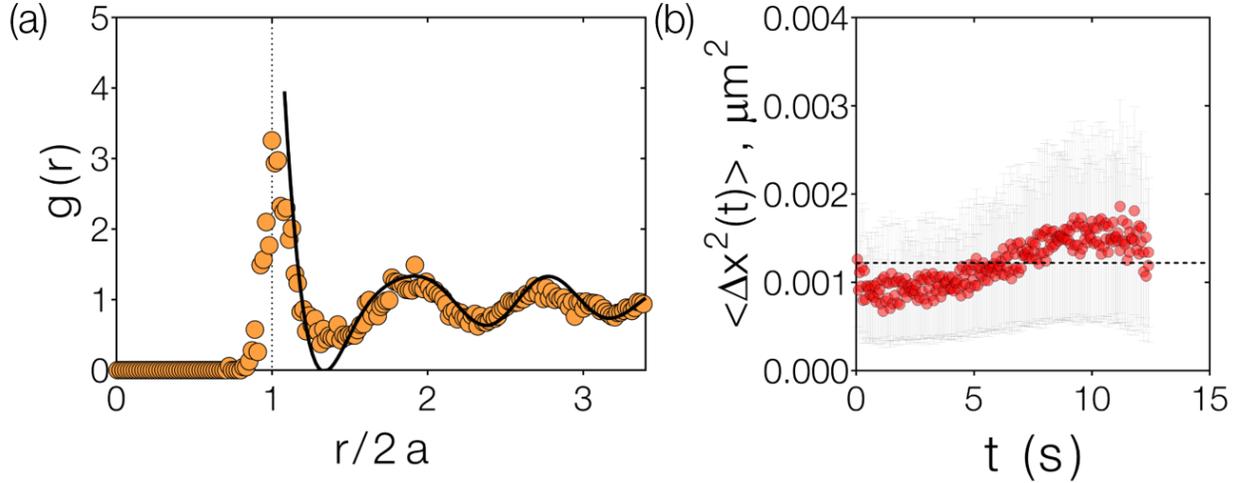

**Figure S1** | (a) The filled circles represent the radial distribution of the smooth colloidal suspension at $\phi = 0.56$. The dark line corresponds to the radial distribution function from classical fluid theory with Percus-Yevick closure that mimics hard-sphere microstructure. The matching of peaks in experimental and theoretical radial distributions show that suspensions retain overall hard-sphere like behavior. (b) Mean-squared displacement of a photopolymerized sample to indicate the noise floor of the confocal rheometer setup.

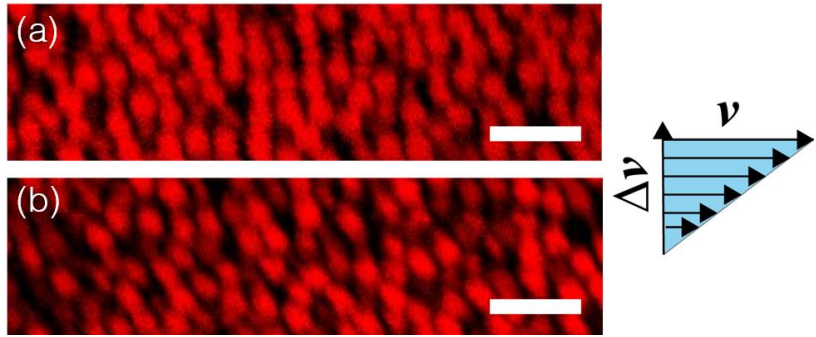

Figure S4 | The z projection of the arrested dynamic microstructure in suspensions of (a) smooth and (b) rough colloids at $\Delta\phi/\phi_{max} \approx 0.08$ and $\beta \approx 0.95$. Shear direction is shown in the inset and the contact networks are aligned in the compressive axis of the shear. Scale = 5 μm.

Sample z-projections of the dynamic microstructure in suspensions of smooth and rough colloids is shown in Figure S4. Centroid identification from similar arrested 3D image stacks enable us generate the respective dynamic contact networks by defining the contact criterion as $2a_{eff}$.

**S.5. Details of the DPD simulations:**

Simulations are done *via* the HOOMD-Blue simulation toolkit where the roughness of different particles is modeled by randomly placing asperities with their centers located on the periphery of the base particle. The base particles are modeled as near hard spheres covered by these smaller patchy asperities, and in our setting we choose the asperity radius to be $0.1a$, where $a$ is the radii of the base particles [5]. The main simulations are based upon a core-modified Dissipative Particle Dynamics (DPD) algorithm [6,7], previously shown to preserve the essential hydrodynamics in dense suspension systems. Over the past two decades, there has been a plethora of research topics that employ DPD as the primary simulation tool [8-10]. DPD is a discrete fluid model, where all particles including the background fluid particles are modeled explicitly and through pairwise interactions. We write the equations of motion for DPD as:

$$m_i \frac{dv_i}{dt} = \sum_{i, i \neq j}^{N_p} \left( F_{ij}^{\ C} + F_{ij}^{\ D} + F_{ij}^{\ R} + F_{ij}^{\ R'} + F_{ij}^{\ H} \right)...........(3)$$

$$\omega_{ij} = (1 - r_{ij} / r_c)....................................................(4)$$

$$F_{ij}^{\ R} = \sigma_{ij} \omega_{ij}(r_{ij}) \Theta_{ij} \Delta t^{-0.5} e_{ij}....................................(5)$$

$$F_{ij}^{\ D} = \gamma_{ij} \omega_{ij}^{\ 2}(r_{ij})(v_{ij}.e_{ij}) e_{ij}....................................(6)$$

$$F_{ij}^{\ C} = \alpha_{ij} \omega_{ij}(r_{ij}) e_{ij}..................................................(7)$$

$$F_{ij}^{\ R'} = F_0^{\ R'} e_{ij}.........................................................(8)$$

$$F_{ij}^{\ H} = \mu_{ij}(v_{ij}.e_{ij}) e_{ij}................................................(9)$$

The solvent particles in DPD are modeled as soft particles and interact through the first three terms on the RHS of Eq. 3. In equation $F_{ij}^{\ C}$, $F_{ij}^{\ D}$ and $F_{ij}^{\ R}$ are the pairwise conservative, dissipative, random forces, respectively. Random and dissipative forces together form the canonical ensemble and satisfy the fluctuation-dissipation requirements [11,12]. The random force introduces thermal fluctuations via a random function, $\Theta_{ij}$. This heat is dissipated by the dissipative force acting against the relative motion of particles $v_{ij} = v_i - v_j$ and $\gamma_{ij}$ is the strength of dissipation, which is coupled with the thermal noise, $\sigma_{ij}$. Together these parameters define the dimensionless temperature as $k_B T = \sigma_{ij}^{\ 2} / \gamma_{ij}$ and $e_{ij}$ is the unit vector. Conservative force, defines the chemical identity of a particle based on its chemical potential/solubility in the system, through a parameter, $\alpha_{ij}$. $F_0^{\ R'}$ is the repulsive force that prevents the colloidal particles to overlap where $F_0^{\ R'}$ is the repulsion constant and $F_0^{\ R'} = 500 k_B T / a$ is used for the base-base interactions. Finally, $F_H^{\ ij}$ is the short-range lubrication force that dissipates the motion of particles through a pair drag term, $\mu_{ij} = 3\pi\mu_0 a_1 a_2 / 2h_{ij}$ based on normal squeezing mode of lubrication. This force enables us to capture the short-range hydrodynamics between the particles, where $a_1$ and $a_2$ are the radii of the interacting colloids and $\mu_0$ is the viscosity of the suspending fluid, and $h_{ij}$ is the surface-surface distance between two interacting colloids. Melrose and Ball [13] have shown previously that the tangential mode of lubrication scaling as $\sim \log(1/h_{ij})$ can be ignored compared to the normal mode of lubrication where stresses scale inversely with the separation distance between the particles. Thus, in this platform, we only consider this normal model of lubrication, squeeze mode, and the tangential motion and hindrance to it naturally comes from the normal interactions between the asperities. We have shown previously that this scheme recovers the same hydrodynamics as a fully resolved Stokesian Dynamics with a complete solution of hydrodynamics for the asperities [14,15]. The hydrodynamic interactions are solved on the asperity-asperity and base-base interactions and not on the asperity-base interaction, because the lubrication forces of high particle size ratio are much weaker than the same size ones [15].

## S.6. Change in suspension contact distribution with increasing shear:

The frictional contacts (when contact criterion is set to $2a_{eff}$) for suspensions containing both smooth (S) and rough (RK) colloids proliferate with increasing applied stress when the suspensions transition from a low-shear viscosity ($\sigma_{\beta=0}$) to a shear thickening regime ($\sigma_{\beta}$), as shown in Figure S5. This transition is accompanied by changes in microstructural anisotropy as particles align along the compressive axis of flow (Figure S4). Here we quantify the change in $<z>$ alone. The change in probability distributions of frictional $<z>$ for suspensions of both smooth and rough particles between experiments and simulations qualitatively matches with the increasing trend. In both the cases there is an increase in frictional contacts with increasing applied $\sigma$. Smooth colloids looses particles of $<z> \approx 2$ and gain particles with $<z> \approx 4$ to 5. On the other hand, rough particles loosed atmost one contact and gain $<z> \approx 2$ to 4. Discrepancies between experiments and simulation may arise from factors such as size polydispersity, surface roughness variability, and the errors associated with the correct centroid determination from image-processing routines.

## S.7. Statistical testing:

Statistical significance of the contact scaling model fit, $\Delta z \sim \Delta \phi^{0.95 \pm 0.07}$, is conducted using reduced chi-squared parameter ($\bar{\chi}_0^2$) defined as:

$$\bar{\chi}_0^2 = \frac{1}{N} \sum_{i=1}^{N} \frac{\left(y_i - f(x_i)\right)^2}{\sigma_i^2} \dots\dots\dots(10)$$

Here, $y_i$ denotes the experimental data, $f(x_i)$ corresponds to the linear fit model, and $\sigma_i$ is the standard deviation of the respective $y_i$ values. In our case the number of data points, $N = 32$ and we obtain a $\bar{\chi}_0^2$ value of 2.12 which is in the $O(10^0)$ that denotes that the fit between experimental observations and corresponding power-law is statistically significant. Deviations are attributed to the experimental uncertainties in obtaining the exact $\phi$ and the estimated contact at shear-jammed state $z_J$.

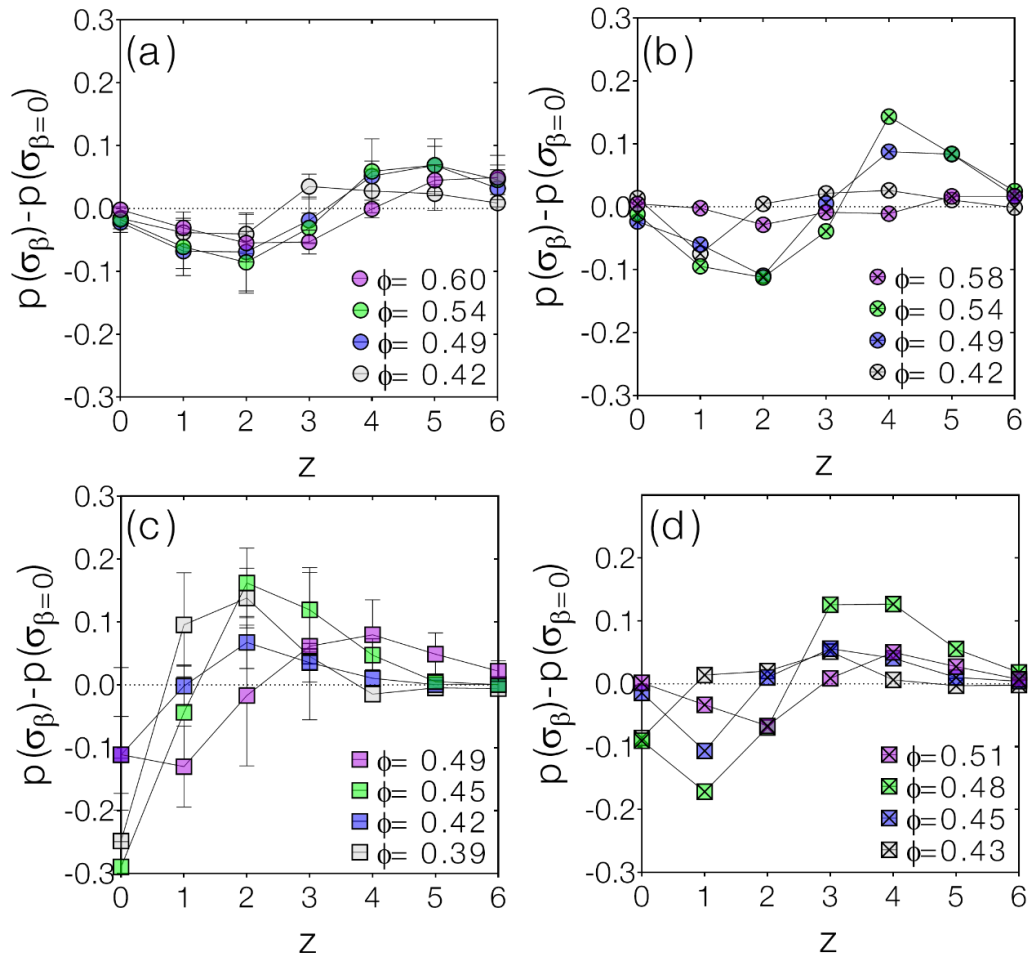

**Figure S5** | The change in stress-dependent contact distribution for suspensions of (a, b) smooth and (c, d) rough colloids in (a, c) experiments and (b, d) simulations.

**S.8. Contact criterion model for shear thickening suspensions:**

A common theory that explains shear thickening mechanics is the transition of particle interaction form a lubrication to a frictional mode above the onset stress, $\sigma^*$. Moreover, the mean-field theoretical description of the shear thickening proposed in the WC model suggests that as the suspension shear thickens, the maximum possible flowable suspension $\phi$ reduces. This is counterintuitive as suspensions on average gains contacts to transmit stress through them by forming force chains. The premise raises the need to define "contact", especially in the case of colloidal suspensions, since they do not actually make an interparticle contact due to factors such as steric brush layer, electrostatics, and solvent lubrication film. At the same time colloids exert a suspension stress, even in the quiescent state, due to the hydrodynamics induced by the Brownian motion, and essentially transmits a macroscopic stress even when no space-spanning force chains are formed. Here, we propose a contact criterion model based on the physics behind the flow mechanics of shear thickening. We define a stress-dependent contact lengthscale where the average contact number $<z>$ can be computed through a mean-field description using function $\mathbb{F}$. Using mechanistic considerations, the nearest neighbor contact is estimated as weighted-average value of the hydrodynamic and frictional contacts as:

$$< z > = < z >_h (1 - \mathbb{F}) + < z >_f \mathbb{F} \dots\dots\dots(11)$$

Here, $<z>_h$ is the hydrodynamic contact which is obtained at low-shear viscosity $\tilde{\sigma}_{\beta=0}$ using a contact criterion that incorporates lengthscales associated with surface roughness, steric polymer brush, and associated polydispersities ($\Delta$) [16], while $<z>_f$ is obtained at $\phi_{J,f}$ with contact criterion set as the average particle diameter. Here, $\mathbb{F}$ is an increasing function of the applied stress $\sigma$ and $\mathbb{F} \in [0,1]$. We do not have enough data to claim the function form of $\mathbb{F}$. In our model, we define the contact criterion ($r'$) as: $r' = r/2a_{eff}$ where $r = r(\mathbb{F})$ and $2a_{eff}$ is the effective average particle diameter. The $\sigma$-dependent contact lengthscale $r$ is defined as:

$$r = 2(r_{\text{surface roughness}} + r_{\text{steric brush}}) + \Delta, \text{ when } \sigma < \sigma^* \ (\mathbb{F} = 0)\dots\dots(12)$$

$$r = 2a_{eff}, \text{ when } \sigma > \sigma^* \ (0 < \mathbb{F} \leq 1)\dots\dots\dots\dots\dots\dots\dots (13)$$

Due to experimental constraints, we do not have access to $<z>_f$ and thus we use equation (11) to obtain $f$ at $\tilde{\sigma}_\beta$. Experimental parameters for the two sets of particles (S- and RK-type) are summarized below in Table S1.

**Table S1.** Input parameters* for the contact model

| Parameter | Smooth (S) | Rough (RK) |
|:---:|:---:|:---:|
| $\phi_{max}$ | 0.64 | 0.54 |
| $<z>_{max}$ | 6.01 | 3.88 |
| $\phi_{J,\beta}$ | 0.61 | 0.51 |
| $<z>_{J,\beta}$ | 4.95 | 3.25 |
| $\phi_{J,f}$ | 0.58 | 0.49 |

*only average parameter used for model validation

Using the average parameter values in Table S1, we estimate $f$ as 0.5 and 0.6 for suspensions of S and RK particles at $\tilde{\sigma}_\beta$, respectively. To simplify the calculations, we assume $\mathbb{F} \approx f$, use equation (11), and the data available in Table S1 to estimate $<z>_f$ as 3.9 and 2.8 for the suspensions of smooth and rough particles, respectively. Using the proposed model, these estimations reveal the microstructural change from a lubricated ($\sigma < \sigma^*$, $\mathbb{F} = 0$, $\phi_{max} \approx 0.64$, and $<z>_{max} \approx 6$) to a frictional state ($\sigma > \sigma^*$, $\mathbb{F} = 1$, $\phi_{J,f} \approx 0.58$, and $<z>_f \approx 4$).

## S.9. Axial forces in shear thickening suspensions

The surface tension of the suspension is important in the computation of the first normal stresses ($N_1$) using a cone-and-plate geometry. Although surface tension effects are usually negligible with standard fluids, dilatant suspensions may contain particles jammed at the surface, which alter the meniscus curvature and significantly decrease $N_1$ [17]. Since the shape of the meniscus was not monitored in this study, we report only the axial force output from the steady shear measurements in Figure. S6. The data show that the competing effects of dilatancy and surface tension are especially apparent for VR and RK colloids at the highest volume fractions ($\phi \geq 0.46$).

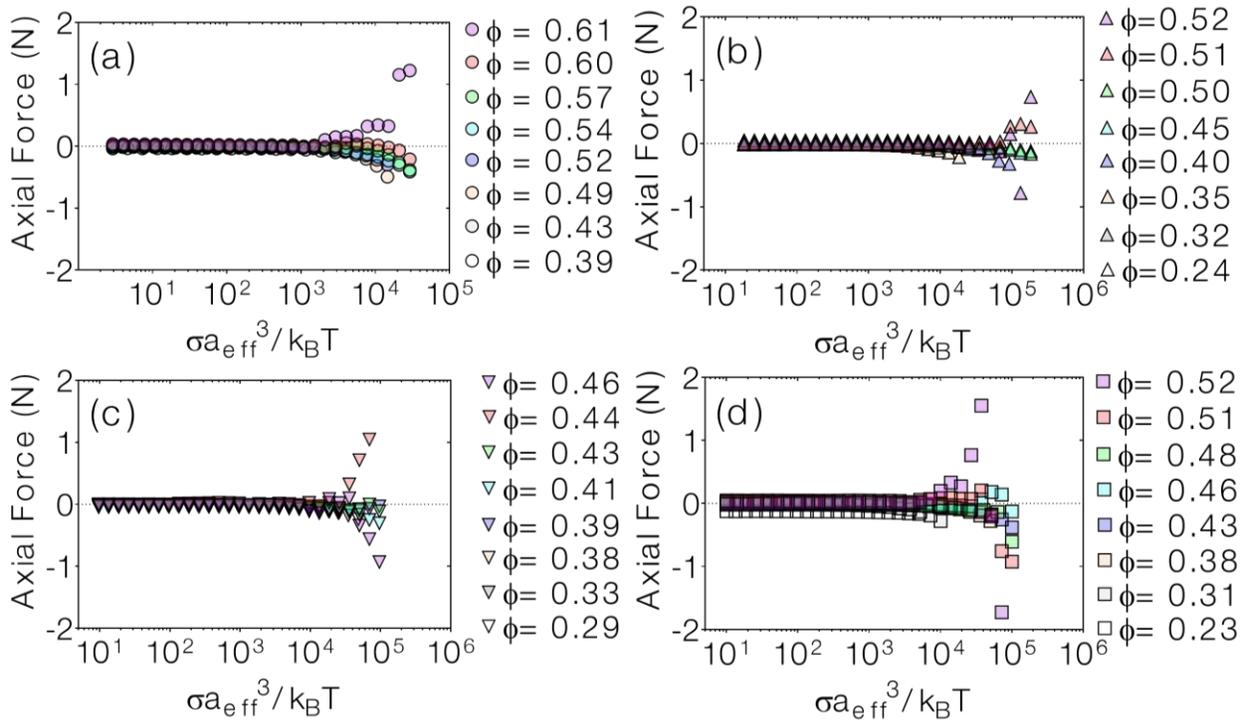

**Figure S2 |** Axial force measurements for all colloidal suspensions tested in this study: (a) S, (b) SR, (c) VR, and (d) RK as a function of shear stress and $\phi$.